\documentclass[doublecol,figures]{epl2} 

\usepackage{graphicx}
\usepackage{dcolumn}
\usepackage{verbatim}
\usepackage{bm}
\usepackage{color,mathrsfs}

\newcommand{\AddrFreiburg}{
 Physikalisches Institut, Albert-Ludwigs-Universit\"at Freiburg - Fakult\"at f\"ur Mathematik und Physik,\\
 D-79104 Freiburg, Germany
 \medskip
 }
\newcommand{\SM}{standard model}
\newcommand{\sinlep}{sin$^2 \theta_{\scriptsize \mbox{eff}}$}

\newcommand{\sefflep}{s$^{2,\mbox{\scriptsize lept}}_{\scriptsize \mbox{eff}}$}
\newcommand{\seffhad}{s$^{2,\mbox{\scriptsize hadr}}_{\scriptsize \mbox{eff}}$}
\newcommand{\seff}{s$^{2}_{\scriptsize \mbox{eff}}$}

\begin{document}

\normalsize
\title{Precision tests of unitarity in leptonic mixing}

\author{L.~Basso\inst{1} \and O.~Fischer\inst{2} \and J.~J.~van~der~Bij\inst{3} }
\institute{ \AddrFreiburg \\
	\inst{1}{lorenzo.basso@physik.uni-freiburg.de\\}
	\inst{2}{oliver.fischer@physik.uni-freiburg.de\\}
	\inst{3}{vdbij@physik.uni-freiburg.de}
}

\date{\today}

\abstract{
In the light of the recent LHC data, we study precision tests sensitive to the violation of lepton universality,
in particular the violation of unitarity in neutrino mixing. Keeping all data we find no satisfatory fit,
even allowing for violations of unitarity in neutrino mixing. Leaving out \sinlep\, from the hadronic forward-backward asymmetry
at LEP, we find a good fit to the data with some evidence of lepton universality violation
at the $\mathcal{O}(10^{-3})$ level.}

\maketitle

The interpretation of the newly discovered scalar boson at the LHC as the Higgs boson 
of the standard model allows for concrete theory predictions for electroweak precision observables. The question arises, whether the measured value for the Higgs mass results in a full agreement between the \SM\, prediction and the experimental observations in the electroweak sector. 
Now that the Higgs boson mass is known, one can look for more subtle deviations from standard model predictions as before.
Although theory and experiment are in rough agreement, some discrepancies remain. There is for example the $2\sigma$ deviation of the $Z$ boson invisible width or the $1.33\sigma$ deviation for the $W$ boson mass $M_W$ \cite{Ferroglia:2012ir}.
More serious is the situation regarding the value of \sinlep, which we will call \seff\, in the following.
There is a large discrepancy between the value from leptonic measurements \sefflep\, and the one based on hadrons
\seffhad.
Only the average of the leptonically and hadronically inferred values agrees with the theory prediction,
but both individual values are off by $\sim 2\sigma$.

The degree of accuracy of any theoretical prediction is related to the precision of the experimental measurements used as
input to the predictions. Typically one takes the most precisely known quantities as input.
For the electroweak theory these are the $Z$ boson mass, the electromagnetic fine-structure constant $\alpha$ and the Fermi constant.
The latter is determined by the muon decay constant, measured from the muon lifetime.
It was shown in Refs.~\cite{Loinaz:2002ep,Loinaz:2004qc} that a violation of unitarity in the mixing among neutrinos modifies
the relation between the muon decay
constant and the Fermi constant (see Eq.~(\ref{eq-Gmu})). Since the Fermi constant is an input to the theory predictions, this modification affects all predictions for the electroweak precision observables.

The latter references pursued this observation further and investigated to which extent a violation of unitary mixing in the neutrino sector would improve the fit to the electroweak precision observables, whereby also constraints from lepton universality
were considered. Oblique corrections were included as free parameters in the analysis, which was necessary at the time because of the lack of knowledge of the Higgs boson mass.  This analysis showed that a large mixing of neutrinos with heavy sterile states
is preferred by the data, however combined with oblique corrections coming from a Higgs boson with a mass in the range of
 several hundreds of GeV.
With the presently known Higgs mass, the large (absolute) values for the oblique corrections advocated above
are not possible anymore. Hence the degree of agreement between theory and experiment is unsatisfactory.
However the analysis relied on old data, that have been improved in the meantime.

When one limits oneself to the electroweak precision observables only, as in Ref.~\cite{Akhmedov:2013hec}, the degree of unitarity violation present in the type-I see-saw \cite{Minkowski:1977sc} with TeV scale right-handed neutrinos can improve the fit to the electroweak precision observables. This analysis allowed for the effect of rather large oblique corrections due to the heavy neutrinos. However the best  fit point is for a configuration in which the contribution of the latter is negligible with respect to the contribution from tree-level effects due to non-unitary mixing. We were able to confirm this result by fitting the same observables with and without the $T$ parameter.
This result then stands in contradiction to the fit of the lepton universality data, when performed separately. We therefore conclude, that the fit of electroweak parameters and lepton universality have to be performed together, as was done in Ref.~\cite{Loinaz:2004qc}, while otherwise misleading interpretations may arise.

In this paper we update the fit of Ref.~\cite{Loinaz:2004qc} in several ways. We use more recent data for lepton universality data and electroweak
measurements. The fact that the Higgs mass is relatively light suggests that it is not possible to achieve large values for the oblique parameters. 
As argued above we expect that also the contribution of heavy neutrino states to the latter is negligible.  
Therefore we perform a $\chi^2$ analysis to check for non-unitary mixing in the neutrino sector, using the most precise experimental precision data, both with and without the contribution from oblique corrections.

The violation of unitarity in the leptonic sector is assumed to be caused by new physics mixing with the \SM\, neutrinos, thereby causing departure from unitarity in the $PMNS$ matrix while keeping the unitarity of the full neutrino mixing matrix.
We consider only the most precise low energy observables for lepton universality, and the set of most precise
electroweak  observables sensitive to deviations in the neutrino sector. These are the $W$ boson mass $M_W$, the invisible decay width of the $Z$ boson $\Gamma_{inv}$, its leptonic decay width $\Gamma_{lept}$ and the weak mixing angle \sinlep.
We believe that these EW observables are more relevant than  the others because they are free from QCD uncertainties, often `biased' by our less accurate knowledge of the hadronic terms.  For  similar reasons, we do not consider NuTeV data, which are 
 shown to be irrelevant in the fit in~\cite{Akhmedov:2013hec}. We discuss critically Ref.~\cite{Ferroglia:2012ir}, where the average of \sefflep\, and \seffhad\, was shown to be consistent with the \SM. We will show that a much better fit of the data is achieved when \seffhad\, is not included. 
This may indicate that this measurement is an accidental statistical outlier or that maybe there is some systematic effect, 
related to the hadronic features of the measurements.

We will find that a global fit can resolve  the discrepancies between theory and experiment in the electroweak precision sector when the stringent tests of lepton universality are considered. 
This is true also when oblique parameters are neglected.
\medskip

We begin our analysis by assuming the $PMNS$ matrix to be non-unitary due to the presence of $n$ additional fermionic fields $N_i$, which are singlets with respect to the \SM~gauge group. For convenience, we call them heavy neutrinos. These new fields mix with the left-handed neutrinos, resulting in the $3+n$ mass eigenstates $\nu_i,\,i=1,...,3+n\,$, of which we conventionally identify the first 3 as the \SM-like neutrinos. We express the mass eigenstates in terms of the flavour basis $\{\nu_{L_\alpha},\,N_i\}$ via a unitary $(3+n)\times(3+n)$ matrix ${\cal U}$, such that:
\begin{equation}
\left(\begin{array}{c} \nu_1 \\ \vdots \\ \nu_{3+n} \end{array}\right) = {\cal U} \left(\begin{array}{c} \nu_{L_e} \\ \vdots \\ N_n\end{array}\right)\,.
\end{equation}
The matrix $\,{\cal U}$ is the generalisation of the leptonic $PMNS$ matrix with rank $3+n$, so that ${\cal U} = U_{PMNS} \Leftrightarrow n=0$. For $n\geq 1$, ${\cal U}_{\alpha\beta}$ ($\alpha,\beta = e,\mu,\tau$) is not unitary. Following~\cite{Loinaz:2002ep,Loinaz:2004qc}, the amount of unitarity violation can be quantified by defining the epsilon parameters
\begin{equation}
\epsilon_\alpha = \sum_{i>3} |{\cal U}_{\alpha i}|^2 = 1- \sum_\beta |{\cal U}_{\alpha \beta}|^2\,.
\label{def-epsilon}
\end{equation}

The recent measurements of the neutrino oscillation parameters bound the epsilon parameters to be less than $\mathcal{O}(10^{-2})$ \cite{Antusch:2008tz}. For the $PMNS$ matrix {\cal U} non-unitary, $\cal{U}{\cal U}^\dagger \neq$ 1. The off-diagonal elements of $\cal{U}{\cal U}^\dagger$ are much more strongly constrained, in particular from  the MEG bound
 on $\mu\to e\gamma$~\cite{Antusch:2008tz}. In the rest of this letter, we will consider the MEG bound to be satisfied and hence neglect the off-diagonal elements, concentrating on the epsilon parameters only.

The epsilon parameters affect the ratio of weak coupling constants $g_e,g_\mu,g_\tau$ of the electron, muon and tau, respectively. They can be experimentally measured in low energy data~\cite{Loinaz:2002ep}:
\begin{eqnarray}\label{low_1}
\left(\frac{g_\mu}{g_e}\right)_\tau^2 & = & \frac{BR_{\tau\to \mu}}{BR_{\tau\to e}} \frac{f[x_{e\tau}^2]}{f[x_{\mu\tau}^2]}\,,\\
\left(\frac{g_\tau}{g_e}\right)_\tau^2 & = & \frac{\tau_\mu}{\tau_\tau} BR_{\tau \to \mu} x_{\mu \tau}^5 \frac{f[x_{e\mu}^2]}{f[x_{\mu\tau}^2]} \frac{\delta_W^\mu \delta_\gamma^\mu}{\delta_W^\tau \delta_\gamma^\tau}\,,\\
\left(\frac{g_\mu}{g_e}\right)_\pi^2 & = & (1 + \delta_{e \mu}^{rad}) \frac{BR_{\pi \to \mu}}{BR_{\pi \to e}} x_{e \mu} \frac{1 - x_{e \pi}^2}{1 - x_{\mu \pi}^2}\,,\\ \label{low_4}
\left(\frac{g_\tau}{g_\mu}\right)_\pi^2 & = & \frac{\tau_\pi}{\tau_\tau} (1 + \delta_{\tau \pi}^{rad}) \frac{BR_{\tau \to \pi}}{BR_{\pi \to \mu}} 2 x_{\mu \tau}^2 x_{\pi \tau} \frac{1 - x_{\mu \pi}^2}{1 - x_{\pi \tau}^2}\,,
\end{eqnarray}
where $BR_{\alpha \to \beta}$ is the branching ratio of particle $\alpha$ into $\beta$, $\tau$ the life time, $x_{\alpha \beta}=m_\alpha/m_\beta$,  $f[x]=1-8x+8x^3-x^4-12x^2\log[x]$ the phase-space factor, $\delta_V^\alpha$ the radiative correction to the decay of particle $\alpha$ caused by the gauge boson $V$ \cite{Ferroglia:2013dga}, and $\delta_{\alpha \beta}^{rad}$ the radiative correction to the ratio of $\alpha$ and $\beta$ decays \cite{Decker:1994ea}. The correlation between the low energy observables can be found in Ref.~\cite{Loinaz:2002ep}. We summarise the experimental values in Table~\ref{precisionobservables}.
The theory prediction for the above formulas is naively 
\begin{equation}
\frac{g_\alpha}{g_\beta} = 1-\frac{\epsilon_\alpha - \epsilon_\beta}{2}.
\end{equation}

Besides directly modifying all the observables in which SM neutrinos are involved, the epsilon parameters  enter into the predictions for all  precision observables by affecting the Fermi constant via the following relation:
\begin{equation}
G_\mu^2 = G_F^2 (1-\epsilon_e)(1-\epsilon_\mu)\,,
\label{eq-Gmu}
\end{equation}
with $G_\mu$ being the Fermi constant as measured in muon decays, and $G_F$ being the actual Fermi parameter.

Since $G_\mu$ is an input parameter, the theory prediction for all the electroweak precision observables is modified.
For this analysis only the most precise measurements should be considered. These are $M_W,\Gamma_{lept},\Gamma_{inv}$, the unitarity of the CKM matrix and \sinlep, which we choose as the set of {\it relevant observables}. Because of the large discrepancy between the measurements of \sinlep\, we keep the leptonic (\sefflep) and hadronic (\seffhad)
measurements as separate data points. The experimental values are listed in Table~\ref{precisionobservables}.

In general, the contribution from heavy neutrinos to oblique parameters is very small. However in Ref.~\cite{Akhmedov:2013hec}
a large value to $\alpha T = \Delta \rho$ due to a cancellation mechanism was found.
Therefore we perform the fit also allowing for this parameter to be non-zero. 

The theory prediction of the {\it relevant} electroweak precision observables can thus be expressed as~\cite{Loinaz:2002ep,Loinaz:2004qc}:
\begin{eqnarray}\label{EW_1}
\frac{M_{W}}{\left[M_{W}\right]_{\mbox{SM}}} &=& 1 + 0.11\,( \epsilon_{e}+\epsilon_{\mu}) \nonumber \\
	& & + 0.0056 T \label{eqn-obs-1}\\
\frac{\Gamma_{\mbox{inv}}/\Gamma_{\mbox{lept}}}{\left[\Gamma_{\mbox{inv}}/\Gamma_{\mbox{lept}}\right]_{\mbox{SM}}}&=&1
-0.76\,(\epsilon_{e}+\epsilon_{\mu})-0.67\,\epsilon_{\tau} \nonumber \\
	& &  -0.0015 T \label{eqn-obs-2} \\
\frac{\Gamma_{\mbox{lept}}}{\left[\Gamma_{\mbox{lept}}\right]_{\mbox{SM}}}&=&1 +0.60\,(\epsilon_{e}+\epsilon_\mu \label{eqn-obs-4})\nonumber \\
	& &  + 0.0093 T\\  \label{EW_4}
\frac{\sin^{2}\theta^{\mbox{lept}}_{\mbox{eff}}}
{\bigl[\sin^{2}\theta^{\mbox{lept}}_{\mbox{eff}}\bigr]_{\mbox{SM}}}&=&1
-0.72\,(\epsilon_{e}+\epsilon_\mu) \nonumber \\
	& & -0.011 T \label{eqn-obs-3}\\ \label{EW_5}
CKM &=& 1+\epsilon_\mu \,.
\end{eqnarray}

As argued above, electroweak observables can not be fit independently from the lepton universality data. Hence, 
we perform a fit of the parameters $\epsilon_e,\epsilon_\mu,\epsilon_\tau$ to the low energy observables (eqs.~(\ref{low_1})--(\ref{low_4})) and to the relevant electroweak observables (eqs.~(\ref{EW_1})--(\ref{EW_5})) in a standard $\chi^2$ analysis. The data presented in Table~\ref{precisionobservables} is taken from Refs.~\cite{Baak:2012kk};
the values of the weak mixing angle are found in Ref.~\cite{Ferroglia:2012ir}.

\begin{table}[h]
\begin{center}
\begin{tabular}{|l|c|c||}
\hline 
Observable & Experiment & \SM \\
\hline \hline 
$(g_\mu/g_e)_\tau$	& 1.0020(16)	& 1.0	 \\
$(g_\tau/g_e)_\tau$	& 1.0029(21)	& 1.0	 \\
$(g_\mu/g_e)_\pi$ 	& 1.0021(16)	& 1.0	 \\
$(g_\tau/g_\mu)_\pi$ 	& 0.9965(33)	& 1.0	 \\
$CKM$	 & 0.9999(6) 	& 1.0 \\
\hline
$M_W$ (GeV) 		& 80.385(15) 	& 80.359(11)  \\
$\Gamma_{\rm inv}/\Gamma_{\rm lept}$ 	& 5.942(16) & 5.9721(2) \\
$\Gamma_{\rm lept}$ (MeV) & 83.984(86) 	& 84.005(15)  \\
\sefflep & 0.23113(21) 	& 0.23150(1)  \\
\seffhad    & 0.23222(27) 	& 0.23150(1) \\
\hline
\end{tabular}
\caption{Experimental results and \SM\, prediction for lepton universality and electroweak observables. The theoretical predictions and experimental values are taken from Refs.~\cite{Baak:2012kk}. The values of the Weinberg angle are taken from Ref.~\cite{Ferroglia:2012ir}.}
\label{precisionobservables}
\end{center}
\end{table}

\begin{table}[h]
\begin{center}
\begin{tabular}{|l|c|c|c|c|}
\hline 
Observable &  $\chi _{SM}^2$ &$\chi_{T}^2$ & $\chi_{\epsilon}^2$ & $\chi_{\epsilon + T}^2$  \\
\hline \hline 
$(g_\mu/g_e)_\tau$	& 19.8 & 18.8 & 17.5 & 17.4 \\
$(g_\tau/g_e)_\tau$	& 20.3 & 19.3 & 14.0 & 13.5 \\
$(g_\mu/g_e)_\pi$ 	& 19.7 & 18.6 & 17.4 & 17.2 \\
$(g_\tau/g_\mu)_\pi$ 	& 20.0 & 19.0 & 17.3 & 17.3 \\
$CKM$	 		& 21.3 & 20.3 & 15.9 & 15.2 \\
\hline
$M_W$ (GeV) 		& 19.4 & 19.4 & 16.9 & 11.6 \\
$\Gamma_{\rm inv}/\Gamma_{\rm lept}$ 	 & 17.8 & 16.9 & 15.8 & 15.4 \\
$\Gamma_{\rm lept}$ (MeV) & 21.4 & 20.2 & 17.6 & 17.5 \\
\sefflep			& 18.2 & 18.1 & 16.2 & 16.0 \\
\seffhad			& 14.2 & 10.5 & 5.3 & 5.3 \\
\hline
\hline 
Total $\chi^2$ 		& 21.3 & 20.3 & 18.0 & 18.0 \\
\hline
\end{tabular}
\caption{The $\chi^2$ for the \SM\, ($\chi^2_{SM}$), the minimum with unitarity violation ($\chi_{\epsilon}^2$) with unitarity violation and the $T$ parameter ($\chi^2_{\epsilon+T}$), and the $T$ parameter only are evaluated {\it excluding} the entry on each line. The total $\chi^2$ (considering all entries) is given for reference.}
\label{chisquared}
\end{center}
\end{table}

In Table~\ref{chisquared} we present the results of the $\chi^2$ analysis.
The first thing to notice is, that if one keeps all data points, the standard model has a bad
$\chi^2=21.3/10$, corresponding to a probability of about 2\%. Adding the epsilon parameters 
does not improve the situation giving a $\chi^2= 18.0/7$, with a probability of about 1\%.
Strictly speaking these data would therefore rule out the standard theory of the electroweak interaction even
allowing for neutrino mixing effects. However it is a fact that the theory has a very consistent structure and that 
no reasonable form of new physics is known that could explain the situation, in particular because the LHC  
has found no new signs of new physics.
Therefore we decided to consider the possibility that one of the measurements
is `wrong', for whatever reason, maybe a statistical fluke or a misunderstanding of systematics.
Therefore we reanalysed the data, leaving out one point at a time. The results are listed in the table.
The only really good fit we find is, when we remove \seffhad\, from the data and allow for the epsilon parameters to be present.
We found a $\chi^2=5.3/6$, corresponding to a probability of about 50\%. One notices that the 
presence of the $T$ parameter has little effect on the goodness of the fit. We will therefore not consider it in the following.
 Without \seffhad\, the standard model has a probability
of about 12\%, however the improvement by allowing for unitarity violation in neutrino mixing is quite large.

The goodness of the fit including the epsilon parameters when \seffhad\, is excluded is the main result of this letter. It shows that there is an indication that \seffhad\, is `wrong' and that a violation of  unitarity in the neutrino mixing matrix is present. 
On the other hand, the inclusion of the $T$ parameter does not play an important role.
From now on, \seffhad\, and the $T$ parameter are no longer considered in the fits. 

The values of the $\chi^2$ and of the epsilon parameters for the best fit point are
\begin{equation}
\chi^2_\epsilon = 5.33 \qquad \begin{array}{c} \epsilon_e = (25.2 \pm 8.1) \times 10^{-4}  \\ \epsilon_\mu = (-3.4 \pm 4.8) \times 10^{-4}\\ \epsilon_\tau = (18.4 \pm 27.6) \times 10^{-4} \end{array}\label{eq-fit}\,,
\end{equation}
with a correlation matrix of
\begin{equation}\label{corr}
\rho = \left( \begin{array}{ccc}
    1 & -0.32 & - 0.12\\
    -0.32  & 1 & - 0.04\\
    - 0.12  & - 0.04  & 1 
    \end{array} \right)\,,
\end{equation}
where the first, second and third row/column correspond to $\epsilon_e,\,\epsilon_\mu$ and $\epsilon_\tau$, respectively. The data suggests that there is unitarity violation in the neutrino sector coming only from the electron flavour, with the remaining two flavours compatible with zero.
However $\epsilon_\tau$ is badly constrained.
The result of our fit respects the experimental bounds from neutrino oscillation experiments, which require $\epsilon < \mathcal{O}(10^{-2})$ \cite{Antusch:2008tz}.

\begin{figure}
\begin{center}
\includegraphics[width=0.4\textwidth]{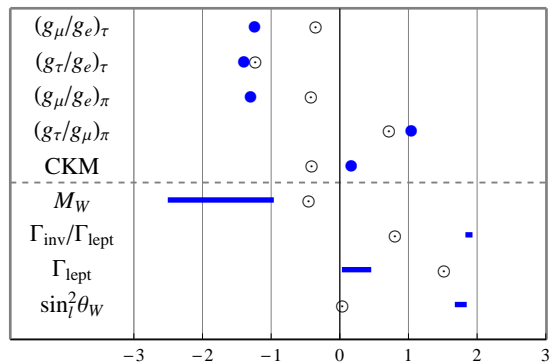}
\end{center}
\caption{Results of the $\chi^2$ analysis. The blue lines represent the \SM\, value and its (theoretical) uncertainty. The circledots denote the best fit results for each observable from the $\chi^2$ analysis. The x-axis scale is set by the respective experimental error, the origin is given by the experimental value.}
\label{fig_fit}
\end{figure}

The best fit point is pictorially shown in Fig.~\ref{fig_fit}. The agreement of the  quantities with the theory prediction is clearly improved with respect to the \SM. The only exception is the leptonic width of the $Z$-boson, off now by $\sim 1.4\sigma$. 
The improvement on all other observables, now off by less than $1\sigma$, compensates for it, which is plain from the value of the total $\chi^2$ at the minimum.

\begin{figure}
\begin{center}
\includegraphics[width=0.3\textwidth]{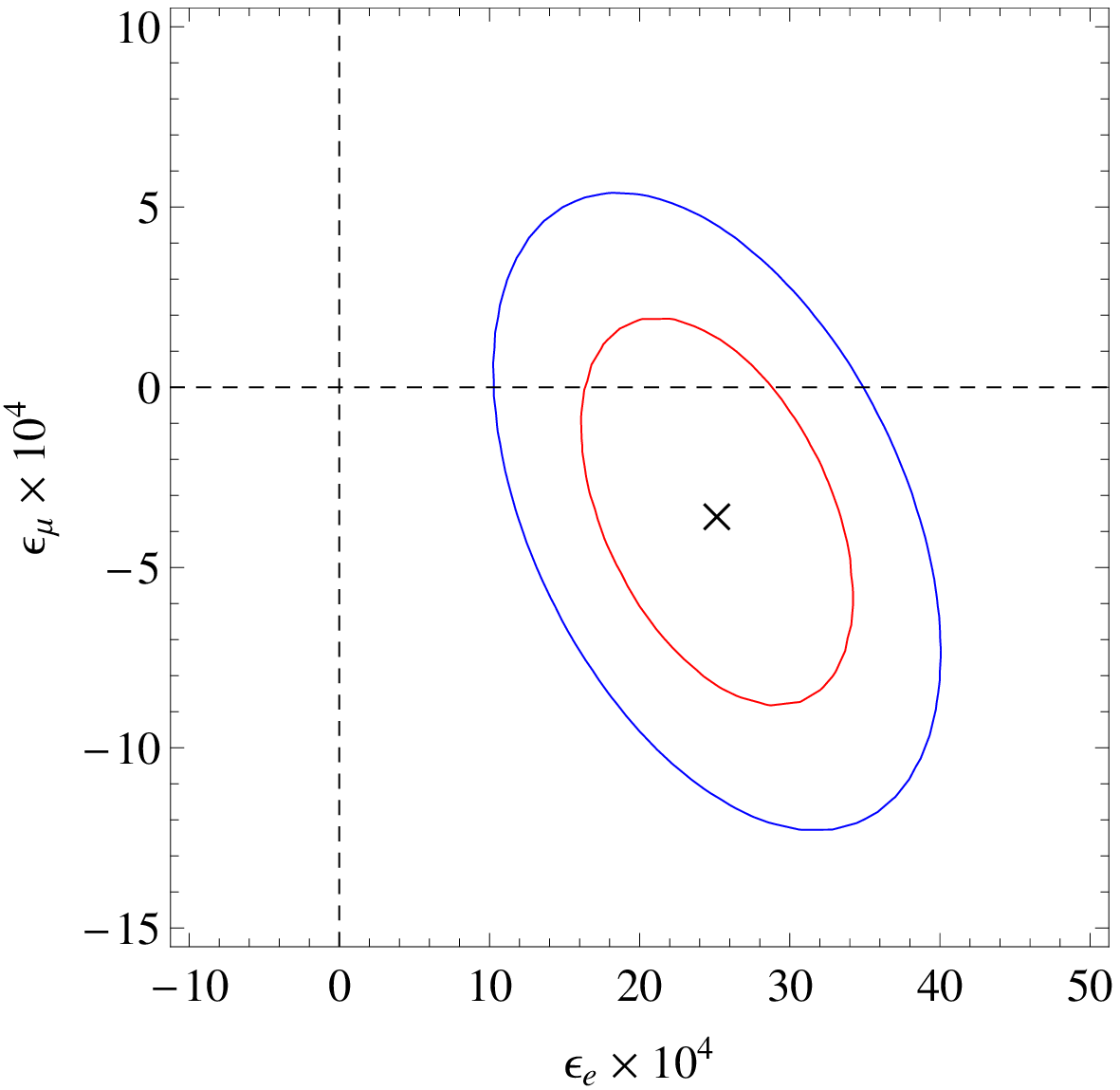}\\
\includegraphics[width=0.3\textwidth]{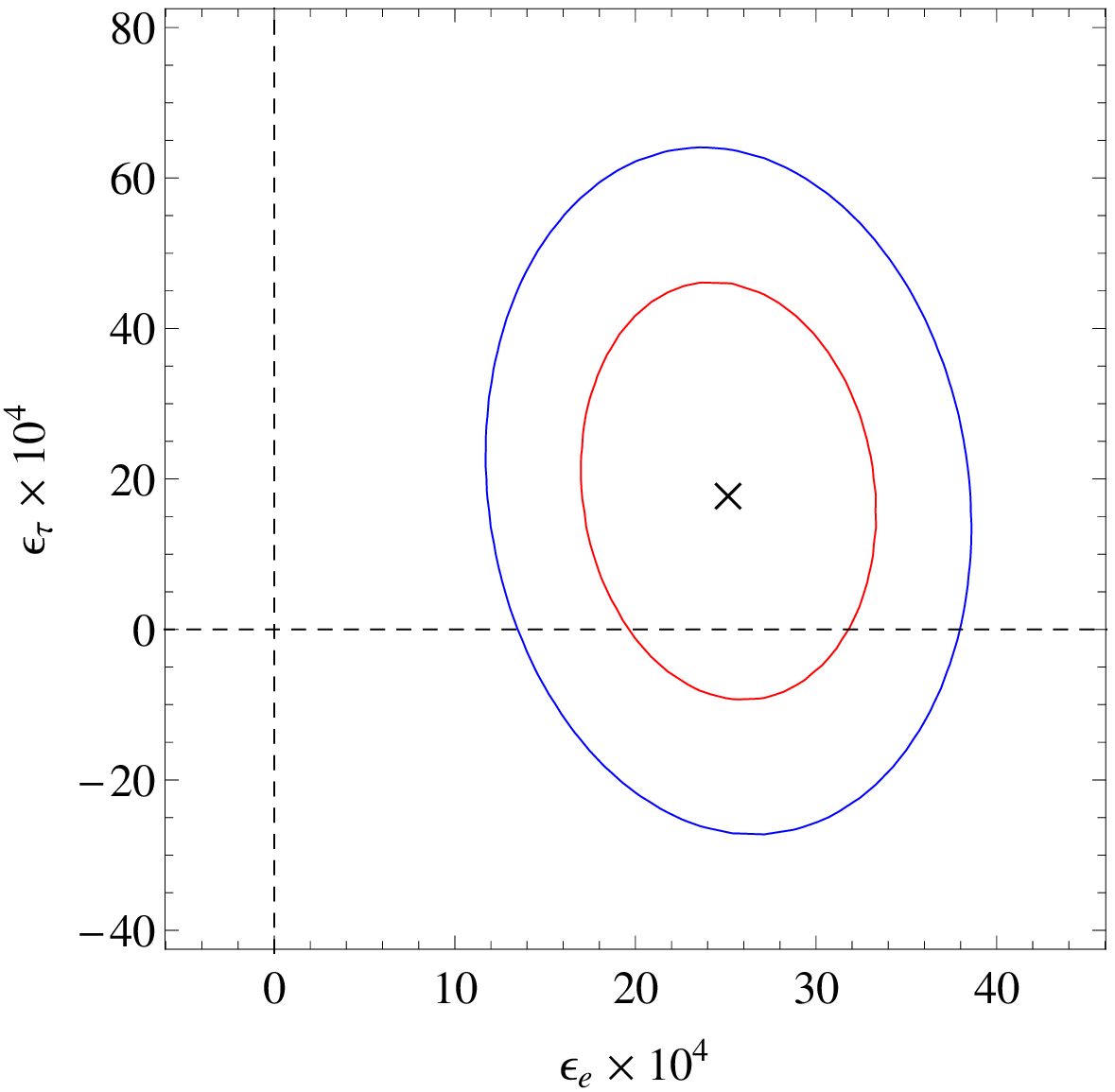}\\
\includegraphics[width=0.3\textwidth]{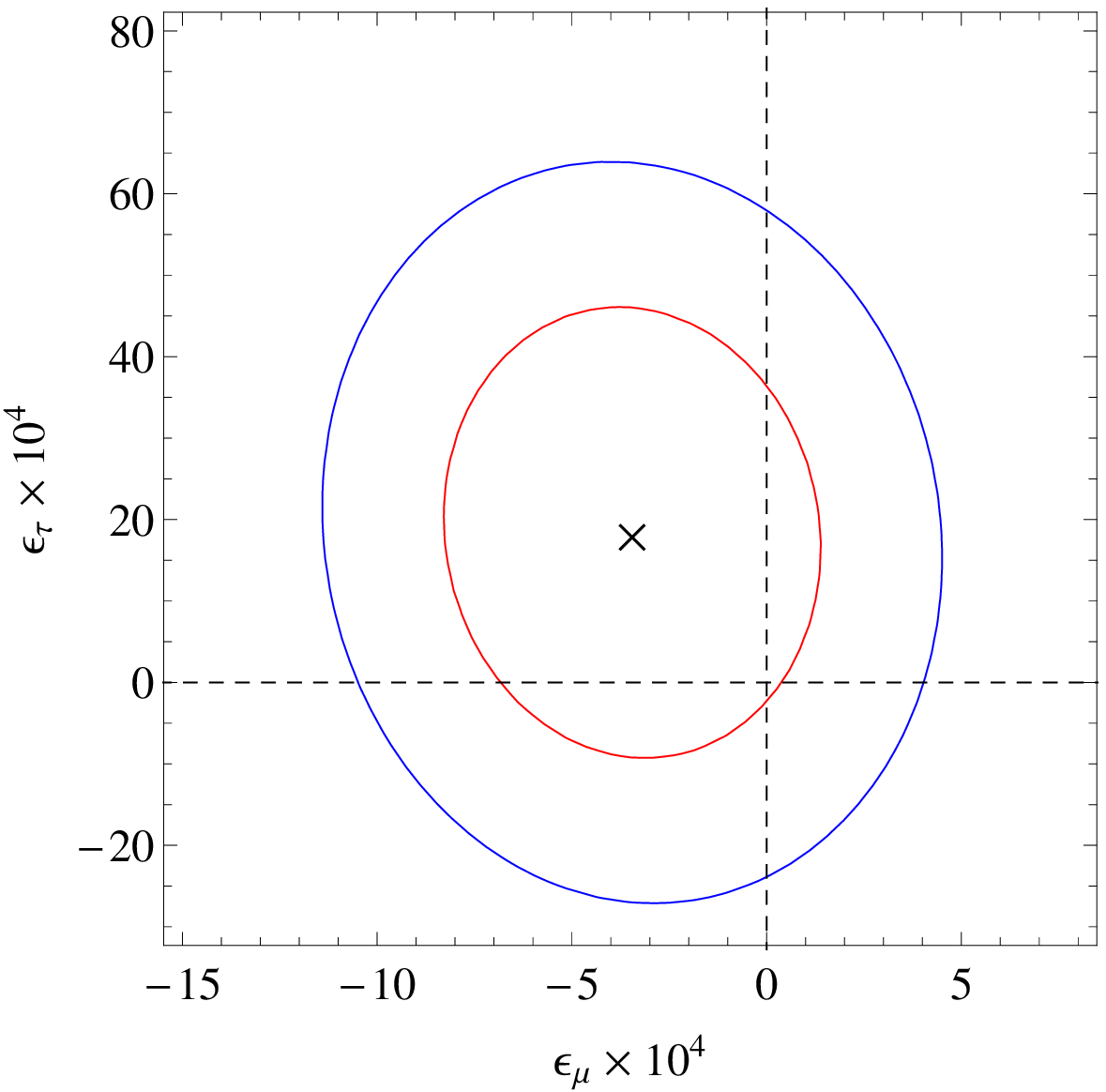}
\end{center}
\caption{Contour plots of $\Delta \chi^2$. The inner line represents the 68\%, the outer one the 90\% confidence level contours. The cross represents the best fit point.}
\label{fig-flatness}
\end{figure}

The presence of extra neutral fermionic fields mixing with neutrinos has also phenomenological implications, which are further subject to experimental constraints. One such constraint comes from the upper limit on the decay rate of $\mu \to e \gamma$, given by the MEG collaboration~\cite{Adam:2013}:
\begin{equation}
{\rm BR}^{\rm MEG}_{\mu^+ \to e^+ \gamma} < 5.7 \cdot 10^{-13}\,.
\label{eq-MEG}
\end{equation}
A conservative theoretical upper bound for the contribution of the heavy neutrinos, as investigated in this letter, is given by (compare Ref.~\cite{Antusch:2008tz}):
\begin{equation}
{\rm BR}^{th}_{\mu\to e \gamma} \leq \frac{3\, n\, \alpha }{8 \pi} \epsilon_e \epsilon_\mu\,,
\end{equation}
where $n$ is the number of additional heavy neutrinos. The values for $\epsilon_e,\epsilon_\mu$ from Eq.~(\ref{eq-fit}), can be consistent with  the constraint in Eq.~(\ref{eq-MEG}), which however indicates that $\epsilon_\mu$ should be quite small,
consistent with zero in the fit to the precision data.

A second constraint comes from the neutrinoless double beta decay \cite{Blennow:2010th,Ibarra:2010xw}, which is given by the EXO collaboration~\cite{Auger:2012ar}:
\begin{equation}
|\left< m_{ee}^{th} \right>|^{\rm EXO} \leq 0.4 \mbox{ eV}\,.
\end{equation}
In assuming conservatively that the contributing heavy neutrino fields have all the same mass $m$, and that all contribute with the same amount to the unitarity violation, as given by Eq.~(\ref{eq-fit}), we get
\begin{equation}\label{0bnn_pred}
|\left< m_{ee}^{th} \right>|^{th} \leq \left| \sum_{i\leq 3} {\cal U}_{e i}^2 m_i + \epsilon_e \frac{q^2}{m} \right|f(A)\,,
\end{equation}
where $q \leq 0.9$ GeV and $f(A)$ is an efficiency factor, which is of order $10^{-2}$ for the material used by the EXO collaboration~\cite{Blennow:2010th}. In Eq.~(\ref{0bnn_pred}) we made use of the unitarity of the neutrino mixing matrix $\mathcal{U}$. We thus obtain a  conservative lower limit on the mass of the heavy neutrinos of ${\cal O}(100)$ TeV.
Of course this limit assumes that the extra neutrinos have Majorana masses. In the case of Dirac masses neutrinoless double-beta decay is absent.

\medskip
From the analysis we can conclude that the combination of the most precise measurements 
is sensitive to the violation of lepton universality at the level of  $\mathcal{O}(10^{-3})$.
We found some evidence ($\sim \,3\,\sigma$) that such a violation, presumably due to a non-unitary mixing among the three standard neutrinos,
is indeed present. We had to leave out the 
LEP-measurement \seffhad\,. Keeping the measurement, no good fit to the data appears possible with or without
the neutrino mixing. Leaving out this measurement and allowing for non-unitary neutrino mixing, a very good fit to the data 
can be found. This is an a posteriori justification not to include this measurement.
Contrary to the conclusions in~\cite{Loinaz:2004qc} we found no need to include additional oblique corrections in the analysis.
We found a picture consistent with the limits from flavour-violating lepton decays, with $\epsilon_e=\mathcal{O}(10^{-3})$,
 $\epsilon_{\mu}$ small
and $\epsilon_{\tau}$ not very well constrained. Though the indications are clear the situation is not quite satisfactory.
However upcoming experiments can improve the situation. First the discrepancy between \seffhad\, and \sefflep\,,
which has affected precision analyses since LEP days, should become clarified by planned  experiments in Mainz and JLAB.
Moreover $\tau$-factories  could be helpful to improve the precision on the branching ratios of the relevant decays. 
Furthermore, an improved measurement of $M_W$ at the LHC, combined with higher order theoretical calculations
will be useful.  An improvement of a factor two in $M_W$ and \seff\,,
assuming the same central values, would lead to a
$\sim \,5.3 \sigma$ effect, sufficient to claim a discovery.

\medskip
This work is supported by the 
Deutsche Forschungsgemeinschaft through the Research Training Group grant
GRK\,1102 \textit{Physics at Hadron Accelerators} and by the
Bundesministerium f\"ur Bildung und Forschung within the F\"order-schwerpunkt
\textit{Elementary Particle Physics}.

\bibliographystyle{unsrt}

\begin{thebibliography}{}

\bibitem{Ferroglia:2012ir}
  A.~Ferroglia and A.~Sirlin,
  Phys.\ Rev.\ D {\bf 87} (2013) 037501
  [arXiv:1211.1864 [hep-ph]].


\bibitem{Loinaz:2004qc}
  W.~Loinaz, N.~Okamura, S.~Rayyan, T.~Takeuchi and L.~C.~R.~Wijewardhana,
  Phys.\ Rev.\ D {\bf 70} (2004) 113004
  [hep-ph/0403306].

\bibitem{Loinaz:2002ep}
  W.~Loinaz, N.~Okamura, T.~Takeuchi and L.~C.~R.~Wijewardhana,
  Phys.\ Rev.\ D {\bf 67} (2003) 073012
  [hep-ph/0210193].

\bibitem{Akhmedov:2013hec} 
  E.~Akhmedov, A.~Kartavtsev, M.~Lindner, L.~Michaels and J.~Smirnov,
  JHEP {\bf 1305}, 081 (2013)
  [arXiv:1302.1872 [hep-ph]].

\bibitem{Minkowski:1977sc}
  P.~Minkowski,
  Phys.\ Lett.\ B {\bf 67} (1977) 421.\\
  T.~Yanagida,
  Conf.\ Proc.\ C {\bf 7902131} (1979) 95.\\
  M.~Gell-Mann, P.~Ramond and R.~Slansky,
  Conf.\ Proc.\ C {\bf 790927} (1979) 315
  [arXiv:1306.4669 [hep-th]].\\
  R.~N.~Mohapatra and G.~Senjanovic,
  Phys.\ Rev.\ D {\bf 23} (1981) 165.

\bibitem{Antusch:2008tz}
  S.~Antusch, J.~P.~Baumann and E.~Fernandez-Martinez,
  Nucl.\ Phys.\ B {\bf 810} (2009) 369
  [arXiv:0807.1003 [hep-ph]].\\
  S.~Antusch, C.~Biggio, E.~Fernandez-Martinez, M.~B.~Gavela and J.~Lopez-Pavon,
  JHEP {\bf 0610}, 084 (2006)
  [hep-ph/0607020].\\
  T.~Ohlsson,
  Rept.\ Prog.\ Phys.\  {\bf 76}, 044201 (2013)
  [arXiv:1209.2710 [hep-ph]].

\bibitem{Ferroglia:2013dga}
  A.~Ferroglia, C.~Greub, A.~Sirlin and Z.~Zhang,
  Phys.\ Rev.\ D {\bf 88} (2013) 033012
  [arXiv:1307.6900 [hep-ph]].

\bibitem{Decker:1994ea}
  R.~Decker and M.~Finkemeier,
  Nucl.\ Phys.\ B {\bf 438} (1995) 17
  [hep-ph/9403385].\\
  W.~J.~Marciano and A.~Sirlin,
  Phys.\ Rev.\ Lett.\  {\bf 71} (1993) 3629.\\
  R.~Decker and M.~Finkemeier,
  Phys.\ Lett.\ B {\bf 334} (1994) 199.

\bibitem{Baak:2012kk}
  M.~Baak, M.~Goebel, J.~Haller, A.~Hoecker, D.~Kennedy, R.~Kogler, K.~Moenig and M.~Schott {\it et al.},
  Eur.\ Phys.\ J.\ C {\bf 72} (2012) 2205
  [arXiv:1209.2716 [hep-ph]].\\
  J.~Beringer {\it et al.}  [Particle Data Group Collaboration],
  Phys.\ Rev.\ D {\bf 86} (2012) 010001.


\bibitem{Adam:2013}
    J.~Adam {\it et al.}  [MEG Collaboration],
    Phys.\ Rev.\ Lett.\  {\bf 110} (2013) 201801.

\bibitem{Blennow:2010th}
  M.~Blennow, E.~Fernandez-Martinez, J.~Lopez-Pavon and J.~Menendez,
  JHEP {\bf 1007} (2010) 096
  [arXiv:1005.3240 [hep-ph]].
\bibitem{Ibarra:2010xw}
  A.~Ibarra, E.~Molinaro and S.~T.~Petcov,
  JHEP {\bf 1009} (2010) 108
  [arXiv:1007.2378 [hep-ph]].

\bibitem{Auger:2012ar}
  M.~Auger {\it et al.}  [EXO Collaboration],
  Phys.\ Rev.\ Lett.\  {\bf 109} (2012) 032505
  [arXiv:1205.5608 [hep-ex]].




\end{thebibliography}
  
\end{document}